 \documentclass[aps,prb,preprint,superscriptaddress,reprint,preprintnumbers,amssymb,longbibliography]{revtex4-1}
\usepackage{amsmath}
\usepackage{amssymb}
\usepackage{braket}
\usepackage{bbold}
\usepackage[version=3]{mhchem}
\usepackage{graphicx}
\usepackage{color}
\usepackage{units}
\usepackage{pifont}
\usepackage[caption=false]{subfig}
\usepackage{multirow}
\usepackage[dvipsnames]{xcolor}
\usepackage{hyperref}
\usepackage[final]{changes}
\usepackage{makecell}

\AtBeginDocument{\usepackage{booktabs}}

\definechangesauthor[name=Stephan Mohr, color=red]{SM}
\definechangesauthor[name={Stephan Mohr}, color=ForestGreen]{LG}
\definechangesauthor[name={Stephan Mohr}, color=blue]{LR}
\definechangesauthor[name={Stephan Mohr}, color=magenta]{MM}
\definechangesauthor[name={done}, color=black]{done}

\begin{document}

\author{Stephan Mohr}
\affiliation{Barcelona Supercomputing Center (BSC), C/ Jordi Girona 29, 08034 Barcelona, Spain}
\email{stephan.mohr@bsc.es}

\author{Marc Eixarch}
\affiliation{Barcelona Supercomputing Center (BSC), C/ Jordi Girona 29, 08034 Barcelona, Spain}

\author{Maximilian Amsler}
\affiliation{Laboratory of Atomic and Solid State Physics, Cornell University, Ithaca, New York 14853, USA}

\author{Mervi J. Mantsinen}
\affiliation{Barcelona Supercomputing Center (BSC), C/ Jordi Girona 29, 08034 Barcelona, Spain}
\affiliation{ICREA, Pg.\ Llu\'is Companys 23, 08010 Barcelona, Spain}

% \author{Chu Chun Fu??}
% \affiliation{to be added}

% \author{Mar\'ia Jos\'e Caturla}
% \affiliation{Departamento de F\'isica Aplicada, Facultad de Ciencias, Universidad de Alicante, Alicante, Spain}

\author{Luigi Genovese}
\affiliation{Univ.\ Grenoble Alpes, INAC-MEM, L\_Sim, F-38000 Grenoble, France}
\affiliation{CEA, INAC-MEM, L\_Sim, F-38000 Grenoble, France}
\email{luigi.genovese@cea.fr}

\title{Linear scaling DFT calculations for large Tungsten systems using an optimized local basis}

%\begin{document}

\date{\today}

\begin{abstract}
Density Functional Theory (DFT) has become the quasi-standard for ab-initio simulations for a wide range of applications.
While the intrinsic cubic scaling of DFT was for a long time limiting the accessible system size to some hundred atoms, 
the recent progress with respect to linear scaling DFT methods has allowed to tackle problems that are larger by many orders of magnitudes.
However, as these linear scaling methods were developed for insulators,
they cannot, in general, be straightforwardly applied to metals, as a finite temperature is 
needed to ensure locality of the density matrix.
In this paper we show that, once finite electronic temperature is employed, the linear scaling version of the \textsc{BigDFT} code is able
to exploit this locality to provide
a computational treatment that scales linearly with respect to the number of atoms of
a metallic system.
We provide prototype examples based on bulk Tungsten, which plays a 
key role in finding safe and long-lasting materials for Fusion Reactors.
We believe that such an approach might help in opening the path towards
novel approaches for investigating the electronic structure of such materials,
in particular when large supercells are required.
\end{abstract}

% PACS
%\pacs
%{
%}
 
% Title
\maketitle

\section{Introduction}
\label{sec:Introduction}
% \SM{Write here a few words about Tungsten and Fusion Reactor Materials... Could you do this Mar\'ia Jos\'e?}
One of the big challenges towards the use of Fusion as a source of clean and safe energy is the design of appropriate reactor components.
During the past years, Tungsten-based materials have emerged as very promising candidates, thanks to their high melting point, high thermal conductivity, low coefficient of thermal expansion, high sputtering threshold energy, low tritium retention and low neutron activation~\cite{waseem-tungsten-based-2016}.
% Still there remain many unresolved questions, which need to be properly addressed.
% One way to do so is to use atomistic simulations, in particular on the first-principles level that gives the most accurate answers.
These materials will sustain high radiation levels which will produce defects that alter their mechanical and transport properties.
Understanding the nature of these defects, such as their atomic structure as well as electronic and magnetic properties, is fundamental to build predictive models of microstructure evolution under irradiation.
In this quest, first principles calculations have been crucial to elucidate the properties and characteristics of the smallest defects, single self-interstitials and vacancies~\cite{nguyen-manh-self-interstitial-2006,ventelon-ab-initio-2012}.
But up to now, only clusters with just a few defects can be studied with these methods.

Over the past decades Kohn-Sham (KS) Density Functional Theory~\cite{hohenberg-inhomogeneous-1964,kohn-self_consistent-1965} (DFT) has become the most popular method for first principles quantum mechanical calculations thanks to the good balance between precision and efficiency offered by this approach.
In particular, DFT offers a much better scaling with respect to the system size compared to other ab-initio approaches.
More precisely, the computational cost scales as the third power of the number of Kohn-Sham orbitals, i.e.\ it is $\mathcal{O}(N^3)$.

Nevertheless, this inherent cubic scaling still limits the accessible system sizes to some hundred atoms, and routine DFT calculations are usually only done in this regime.
Fortunately, it is still possible to reduce the cubic complexity of DFT calculation by using so-called linear scaling approaches~\cite{goedecker-linear-1999,bowler-O(N)-2012}, meaning that doubling the number of atoms in a system leads to a computation time that is only twice as large.
The foundation of the linear scaling approaches lies in the locality of the density matrix $F(\mathbf{r},\mathbf{r}')$.
The so-called nearsightedness principle~\cite{kohn-density-1996} states that the properties of the density matrix at a point $\mathbf{r}$ depend only on points $\mathbf{r}'$ in a localized region around $\mathbf{r}$,
and indeed it can be shown that the matrix elements $F(\mathbf{r},\mathbf{r}')$ decay exponentially with the distance $|\mathbf{r}-\mathbf{r}'|$ for insulators and metals at finite temperatures~\cite{cloiseaux-energy-1964,kohn-wannier-1973,rehr-wannier-1974,goedecker-decay-1998,ismail-beigi-locality-1999}.
Since the density matrix is enough to completely describe a quantum mechanical system, this allows to eventually reach a DFT algorithm that scales linearly with system size.

These theoretical foundations have led to a large variety of approaches to perform linear scaling DFT calculations~\cite{goedecker-efficient-1994,goedecker-tight-binding-1995,goedecker1995low,sankey-projected-1994,stephan-order-N-1998,yang-direct-1991,yang-a-local-1991,yang-a-density-matrix-1995,li-density-matrix-1993,mcweeny-some-1960,mauri-orbital-1993,ordejon-unconstrained-1993,ordejon-linear-1995,mauri-electronic-structure-1994,kim-total-energy-1995,hierse-order-1994,hernandez-self-consistent-1995};
an overview over linear scaling methods can be found in Refs.~\onlinecite{goedecker-linear-1999,bowler-O(N)-2012}.
Together with the steadily increasing capacity of today's supercomputers, this has led to various DFT codes that exhibit such a linear scaling algorithm, as for instance \textsc{ONETEP}~\cite{skylaris-introducing-2005,haynes-onetep-2006,mostofi-onetep-2007,skylaris-recent-2008}, \textsc{Conquest}~\cite{bowler-practical-2000,bowler-recent-2006,bowler-an-overview-2010}, SIESTA~\cite{soler-the-SIESTA-2002}, \textsc{Quickstep}~\cite{vandevondele-quickstep-2005}, \textsc{OPENMX}~\cite{ozaki-efficient-2005,ozaki-O(N)-2006}
or \textsc{BigDFT}~\cite{genovese-daubechies-2008,mohr-2014-daubechies,mohr-2015-accurate};
an overview of popular electronic structure codes and methods for large scale calculations can be found in Ref.~\onlinecite{ratcliff-2017-challenges}.

However, these various implementations of linear scaling DFT are usually only applied to systems exhibiting a finite gap,
and abundant demonstrations for metallic systems are missing.
%Indeed it is questionable whether the techniques developed for insulating systems could be directly 
%applied to metals~\cite{aarons-perspective-2016}.
Still there are a few examples of reduced scaling methods for metals in the literature~\cite{aarons-perspective-2016}.
With respect to \textsc{ONETEP}, a recent implementation based on a direct free energy minimization technique allows to
perform calculations for metallic systems, thereby reducing the computational workload~\cite{ruiz-serrano-a-variational-2013}.
%but cannot be considered a real $\mathcal{O}(N)$ method due to necessary diagonalization, which scales cubically~\cite{ruiz-serrano-a-variational-2013}.
In \textsc{OPENMX}, there is an implementation of a divide-and-conquer approach within a Krylov subspace that allows to perform linear scaling 
calculations for metals; however the method seems to require a careful tuning before it 
can be applied to large systems~\cite{ozaki-O(N)-2006}.
Furthermore it has been shown that the approach by Suryanarayana et.\ al.~\cite{suryanarayana-coarse-graining-2013}, which calculates the electronic charge density and 
the total energy directly by performing Gauss quadratures over the spectrum of the Hamiltonian and in this way is capable to reach a 
linear complexity, also works for metallic systems~\cite{ponga-linear-2016}.

These pioneering examples demonstrate that, even if their application is possible,
reduced scaling approaches for large metallic systems have to be considered under a different perspective.
As for such systems the number of degrees of freedom is very large and the electronic structure is complicated,
first-principles calculation become a useful tool only when complementing other approaches, like for instance 
force fields, which are unable to provide quantum-mechanical information.
When such kind of information is required, like for example when studying the arrangement of electrons close to a defective region,
investigation techniques like the ones presented above are or utmost importance.

In this paper we report on the capabilities of \textsc{BigDFT} to perform reduced and eventually also linear scaling calculations 
for a metallic system at finite temperature.
Whereas previous publications have highlighted in detail the accuracy, efficiency and linear scaling capabilities of 
this code for systems with a finite HOMO-LUMO gap~\cite{mohr-2014-daubechies,mohr-2015-accurate}, metals have so far not been considered.
In this paper we demonstrate that the basic algorithm of \textsc{BigDFT} remains stable also for systems with vanishing gap and 
thus allows to routinely perform accurate linear scaling calculation for large metallic systems without the need of additional adjustments.
As specific example we have chosen Tungsten due to its relevance for finding safe and long-lasting materials for fusion reactors.

The outline of this paper is as follows:
In Sec.~\ref{sec:Theory} we focus on the theoretical background, with Sec.~\ref{sec:Overview of the algorithm} summarizing the principles of the linear scaling version of \textsc{BigDFT},
and Sec.~\ref{sec:Challenges for metallic systems} discussing how \textsc{BigDFT} can mitigate the challenges arising for metallic systems.
In Sec.~\ref{sec:Results} we then present numerical results, 
with Sec.~\ref{sec:Accuracy of the linear scaling version} demonstrating the precision that we obtain with
the linear scaling version of \textsc{BigDFT} and Sec.~\ref{sec:Performance} showing various performance figures.

\section{Theory}
\label{sec:Theory}

\subsection{Overview of the algorithm}
\label{sec:Overview of the algorithm}
The detailed implementation of the linear scaling algorithm of \textsc{BigDFT} has been presented in detail in Refs.~\onlinecite{mohr-2014-daubechies,mohr-2015-accurate}.
Here we will give a brief overview over the most important concepts.

The central quantity on which the algorithm is based, namely the density matrix, is represented in a separable way via a set of localized and adaptive basis functions $\{\phi_\alpha(\mathbf{r})\}$, from now on also called ``support functions'', as
\begin{equation}
 F(\mathbf{r},\mathbf{r}') = \sum_{\alpha,\beta} \phi_\alpha(\mathbf{r}) K^{\alpha\beta} \phi_\alpha(\mathbf{r}') \,,
 \label{eq:density_matrix_in_terms_of_support_functions_and_kernel}
\end{equation}
where the matrix $\mathbf{K}$ denotes the ``density kernel''.
From this expression we can easily get the electronic charge density as $\rho(\mathbf{r})=F(\mathbf{r},\mathbf{r})$ and use it for the construction of the Kohn-Sham Hamiltonian,
\begin{equation}
 \mathcal{H}[\rho] = -\frac{1}{2}\nabla^2 + \mathcal{V}_{KS}[\rho] + \mathcal{V}_{PSP} \,,
\end{equation}
where the Kohn-Sham potential $\mathcal{V}_{KS}[\rho] = \int \frac{\rho(\mathbf{r}')}{|\mathbf{r}-\mathbf{r}'|} \,\mathrm{d}\mathbf{r}' + \mathcal{V}_{XC}[\rho]$ contains the Hartree and exchange-correlation potential, and $\mathcal{V}_{PSP}$ denotes the pseudopotential~\cite{willand-norm-conserving-2013} that is used to describe the union of the nuclei and core electrons.
%In \textsc{BigDFT} we use for this latter purpose the norm-conserving GTH-HGH pseudopotentials~\cite{hartwigsen-relativistic-1998} and their Krack variants~\cite{krack-pseudopotentials-2005}, possibly enhanced by a nonlinear core correction~\cite{willand-norm-conserving-2013}.
This Hamiltonian operator gives rise to the Hamiltonian matrix $\mathbf{H}$, defined as
\begin{equation}
 H_{\alpha\beta} = \int \phi_\alpha(\mathbf{r}) \mathcal{H}(\mathbf{r}) \phi_\beta(\mathbf{r}) \mathrm{d}\mathbf{r} \,,
\end{equation}
which can then be used to determine a new density kernel; methods to do so will be discussed later.

In order to obtain a linear scaling behavior, it is necessary to employ a set of \emph{localized} support functions that eventually lead to sparse matrices.
What distinguishes \textsc{BigDFT} from other similar approaches is the special set of localized support functions that it uses.
These are expanded in an underlying basis set of Daubechies wavelets~\cite{daubechies-ten-1992} and are optimized in-situ.
Daubechies wavelets offer the outstanding property of being at the same time orthogonal, systematic and exhibiting compact support.
The in-situ optimization,
together with an imposed approximative orthonormality, results in a set of quasi-systematic support functions offering a very high precision.
This allows to work with a \emph{minimal} set of support functions, meaning that only very few functions per atom are necessary to obtain a very high accuracy.
Obviously, this in-situ optimization comes at some cost compared to an approach working with a fixed set of non-optimized support functions.
However, in the latter case we would require a much larger set to obtain the same precision~\cite{mohr-complexity-reduction-I-2017}, and all matrix operations in the subspace of the support functions, whose scaling is cubic in the worst case, would become considerably more costly.
Apart from that, the use of a minimal basis set also has additional advantages, as for instance an easy and accurate fragment identification and associated population analysis for large systems~\cite{mohr-complexity-reduction-I-2017}, which can be used for a reliable effective electrostatic embedding~\cite{mohr-complexity-reduction-II-2017}.

\subsection{Advantages of \textsc{BigDFT} for metallic systems}
\label{sec:Challenges for metallic systems}
DFT calculations for metallic systems are a challenging task.
Due to the non-zero density of states at the Fermi energy, the occupation of the eigenstates around that energy value can easily jump between occupied and empty during the self-consistency cycle, leading to a phenomenon called ``charge sloshing''.
A solution to this problem is to introduce a finite electronic temperature, leading to the grand-canonical extension of DFT as derived by Mermin~\cite{mermin-thermal-1965}.
In such a setup,
the occupations are smoothed out around the Fermi level and the self-consistency cycle becomes more stable.

In the context of linear scaling approaches, the introduction of a finite temperature has the additional advantage that it intensifies the decay properties of the density matrix, as mentioned in Sec.~\ref{sec:Introduction}, and thus justifies the exploitation of the nearsightedness principle.
Nevertheless, linear scaling calculations for metals remain very challenging.
First of all, the used electronic temperatures must not be too large --- otherwise one would change the physics in a too drastic way --- and thus the density matrix decays much slower compared to finite gap systems.
Moreover, the vanishing gap complicates the calculation of the density kernel.
In \textsc{BigDFT}, we use for this task the \textsc{CheSS} library~\cite{mohr-efficient-2017} --- one of the building-blocks
of the \textsc{BigDFT} program suite --- that offers several different solvers.
In the Fermi Operator Expansion (FOE) method~\cite{goedecker-efficient-1994,goedecker-tight-binding-1995}, which is the linear scaling solver available within \textsc{CheSS}, one has to approximate the function that assigns the occupation numbers --- typically the Fermi function --- with Chebyshev polynomials.
Obviously this method is most efficient if the degree of the polynomial expansion is small.
This is the case if first the spectral width of the involved matrices is very small, and second if the Fermi function that must be approximated with the polynomials is smooth.
Unfortunately the latter condition is violated for metals, since --- even when using a small finite temperature --- the Fermi function that must be approximated exhibits a sharp drop at the Fermi energy and we thus have to approximate a rather step-like function.
As a consequence, it is questionable whether the FOE method can be used in practice for calculations with metals.

\begin{figure}
 \includegraphics[width=1.0\columnwidth]{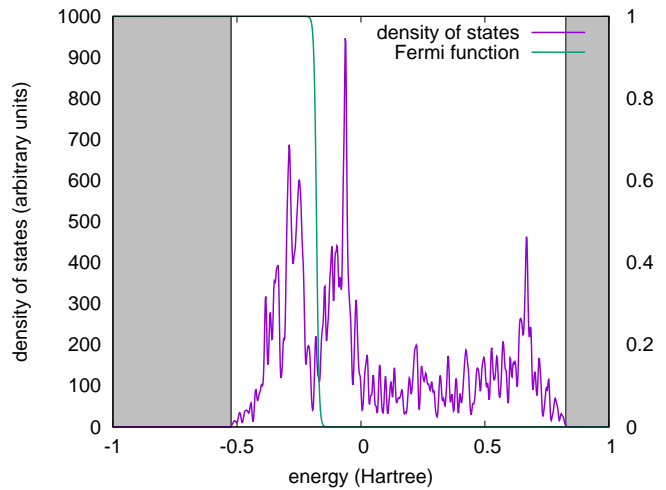}
 \caption{Density of states for the 11x11x11 supercell of bcc Tungsten, containing 2662 atoms.
 The spectral width is very small, which is a direct consequence of the special properties of the support functions used by \textsc{BigDFT}.}
 \label{fig:dos_and_fermifunction}
\end{figure}
Fortunately it turns out that the special properties of the support functions used by \textsc{BigDFT} lead to such a small spectral width that FOE can still be used for metallic systems.
In Fig.~\ref{fig:dos_and_fermifunction} we show the density of states for the 11x11x11 supercell of body centered cubic (bcc) Tungsten, containing 2662 atoms.
As can be seen, the spectral width is even smaller than the default $[-1,1]$ interval for the Chebyshev polynomials.
In this way the polynomial degree required to accurately represent the Fermi function can be kept reasonably small even for metallic systems.

\begin{figure}
 \includegraphics[width=1.0\columnwidth]{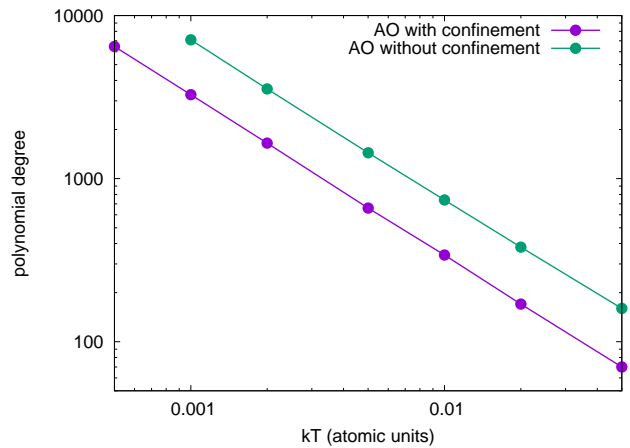}
 \caption{Polynomial degree used by the FOE method within \textsc{CheSS} as a function of the electronic temperature, for two set of atomic orbitals (AO).
 The set that was obtained by solving the atomic Schr\"odinger equation with a confining potential leads to considerably lower values due to its smaller spectral width.}
 \label{fig:npl_as_function_of_temperature}
\end{figure}

This small spectral width is a direct consequence of two concurrent elements.
First of all, the usage of PSP helps by eliminating the need for the treatments of core electrons
whilst smoothening the behavior of the valence KS orbitals close to the nuclei.
The second important point is the special way in 
which \textsc{BigDFT} optimizes the support functions.
As is explained in more detail in Ref.~\onlinecite{mohr-2014-daubechies}, a confining potential is used in order to properly 
localize the support functions during the optimization,
and this confinement also seems to help in reducing the spectral width.
In Fig.~\ref{fig:npl_as_function_of_temperature} we show the polynomial degree used by \textsc{CheSS} as function of the temperature for two sets of atomic orbitals.
Both were obtained by solving the Schr\"odinger equation for the isolated atom within the pseudopotential approach, but in one case we additionally added the confining potential.
In this latter setup, the resulting set of atomic orbitals exhibits a much smaller spectral width (\unit[36.7]{eV}) compared to the case without confinement (\unit[186.4]{eV}), leading to much smaller polynomial degrees.

Nevertheless it is important to note that FOE, even though it can be used very efficiently within \textsc{BigDFT}, is an $\mathcal{O}(N)$ method designed for very large systems.
For intermediate system size, 
alternative solvers within \textsc{CheSS}, such as diagonalization using \textsc{LAPACK}~\cite{anderson-lapack-1999}/\textsc{ScaLAPACK}~\cite{blackford-scalapack-1997} or \textsc{PEXSI}~\cite{lin-accelerating-2013}, might thus be more efficient.
The first method does not exploit the sparsity of the matrices and thus exhibits a cubic scaling, 
whereas the second one scales as $\mathcal{O}(N)$, 
$\mathcal{O}(N^{3/2})$ and $\mathcal{O}(N^2)$ for one-, two- and three-dimensional systems, respectively.
Still, FOE is the method of choice in the limit of very large systems since it is the only solver which scales strictly linearly with system size.

\section{Tests and Considerations}
\label{sec:Results}

In order to demonstrate the accuracy and performance of the linear scaling version of \textsc{BigDFT} for metallic systems, 
we focus on one specific system, namely bcc Tungsten.
All runs were performed using a grid spacing of at most 0.38 atomic units,
the exchange-correlation part was described by the PBE functional~\cite{perdew-generalized-1996}, 
and the Krack HGH pseudopotential~\cite{krack-pseudopotentials-2005} was used.
As we are interested in systems requiring the usage of very large supercells,
we did not considered $k$-points in our calculations.
Nevertheless, we still choose as test-bed for our approach a bulk-like system that can be easily simulated
via $k$-points and small supercells, in order to verify the accuracy of our linear scaling approach.

\subsection{Accuracy of the linear scaling version}
\label{sec:Accuracy of the linear scaling version}

\subsubsection{Energy versus volume}
\label{sec: Energy versus volume}
\begin{figure}
 \includegraphics[width=1.0\columnwidth]{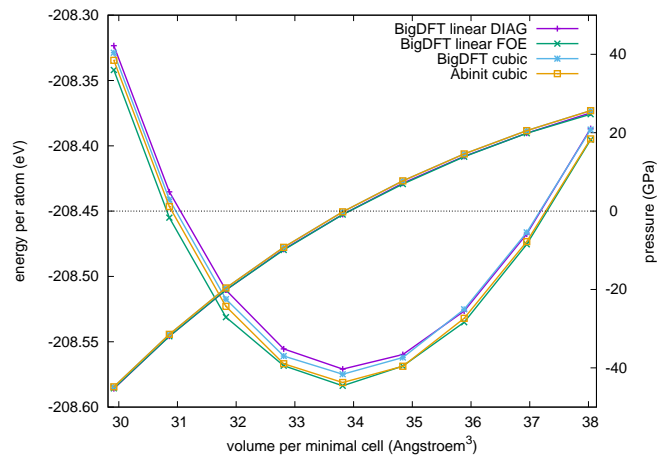}
  \caption{
  Plots of the energy (left axis) and the pressure (right axis) as a function of the cell volume, for the four setups described in Sec.~\ref{sec: Energy versus volume}.
  The linear scaling version of \textsc{BigDFT} yields results that are consistent with those of the two traditional cubic scaling approaches.
  }
  \label{fig:W_energy-vs-volume}
\end{figure}

As a first test we demonstrate that the linear scaling version can accurately calculate the equation of state relating energy and volume.
To this end, we scaled the lattice vectors of the Tungsten system by $\pm4\%$ around its equilibrium value.
We compare four different setups:
(1)/(2) the linear scaling version of \textsc{BigDFT} using a 9x9x9 supercell (containing 1458 atoms) and no $k$-points, using as solver both diagonalization (DIAG) and FOE;
(3) the cubic scaling version of \textsc{BigDFT} using the 2-atoms unit cell and a 9x9x9 $k$-point mesh;
(4) the same setup as (3), but run with the \textsc{Abinit} code~\cite{gonze-first-principles-2002,gonze-a-brief-2005,gonze-abinit-2009,gonze-recent-2016}.
In Fig.~\ref{fig:W_energy-vs-volume} we compare the energy-vs-volume curves for all four setups.
In the same figure we also show, for all four setups, the variation of the pressure as a function of the volume.
%which is related to the derivative of the energy with respect to the volume.
%Therefore, the constant energy offset between the various setups becomes irrelevant and all approaches yield identical results.
As can bee seen from this test, the linear scaling approach correctly determines the optimal lattice parameter.

\subsubsection{Density of states}
As a second test we compare the density of states (DoS) in order to verify that the electronic structure is correctly described.
In Fig.~\ref{fig:W_energy-vs-volume} we compare the DoS of the reference calculation with the cubic version of \textsc{BigDFT} 
and the one obtained with the linear version with diagonalization.
As can be seen, both setups yield an identical DoS for the occupied states.
For the unoccupied ones, the linear version of \textsc{BigDFT} shows some deviations.
However, this is not surprising, since the optimization of the support functions only takes into account the occupied states, 
and a good accuracy can thus only be expected for the latter.
However, as shown in Ref.~\onlinecite{Ratcliff2015Fragment} it is possible to include extra states in the optimization of the support functions in case that the user is interested in low-energy conduction states.
Overall, we see that, thanks to the in-situ optimization of the support functions, the linear scaling version of \textsc{BigDFT} is able to correctly reproduce the electronic structure of a metallic system.
\begin{figure}
 \includegraphics[width=1.0\columnwidth]{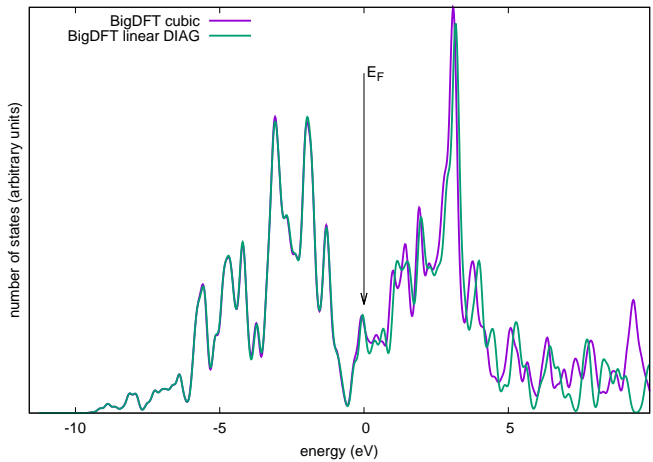}
  \caption{
  Density of states for the cubic scaling version of \textsc{BigDFT} and the linear scaling version with diagonalization.
  The energies were shifted such that the Fermi energy lies at zero.
  Up to the Fermi energy, the linear scaling version of \textsc{BigDFT} yields results that are consistent with those of the traditional cubic scaling approach.
  }
  \label{fig:W_dos_cubic-vs-linear}
\end{figure}

\subsection{Performance}
\label{sec:Performance}

\subsubsection{Scaling with system size}
\begin{figure}
 \includegraphics[width=1.0\columnwidth]{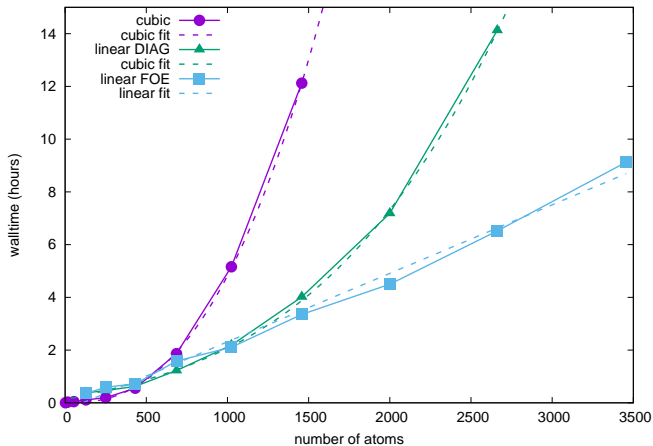}
 \caption{Scaling of the total runtime as a function of the system size, for the cubic approach, the linear approach with diagonalization, and the linear approach with FOE. The calculations were performed at the $\Gamma$ point, and the runs were performed in parallel, using 9600 cores (800 MPI tasks with 12 OpenMP threads).
 }
 \label{fig:scaling_comparison}
\end{figure}

%Now we analyze how the two versions compare in terms of performance.
As anticipated, we expect that DFT calculations of metallic systems at large scales will be
time-consuming compared to similar simulations for insulators.
In Fig.~\ref{fig:scaling_comparison} we show the total runtime as a function of the number of atoms in the system, 
going from the 4x4x4 supercell (128 atoms) up to the 12x12x12 supercell (3456 atoms).
As can be seen, each of the three approaches that we compare --- cubic, 
linear with diagonalization and linear with FOE --- is characterized by a typical system size at which the method 
outperforms the other ones.
For the small systems, the cubic approach is clearly the fastest one.
Above about 500 atoms, the linear approach using diagonalization becomes the method of choice, since the cubic scaling of the diagonalization exhibits a rather small prefactor.
For system sizes beyond about 1000 atoms, however, this cubic scaling starts to become a dominant part, 
and thus the truly linear scaling approach with FOE becomes the fastest option.

\subsubsection{Considerations about the computational cost}
We have demonstrated that the linear scaling version of \textsc{BigDFT} 
can offer an unbiased and unconstrained description of metallic systems
and is thus capable of yielding results 
that are of the same quality as those of a traditional cubic scaling approach.
% therefore by offering the quality of a unbiased, unconstrained description.
However, the inspection of the overall walltime clearly shows that
the application to metals is much heavier compared to insulators.
Considering the CPU-minutes per atom, which can be used as a metric to quantify the computational workload 
for $\mathcal{O}(N)$ codes \cite{mohr-2015-accurate}, and comparing with values obtained for systems such as organic molecules of
light atoms, reveals that the latter run up to two order of magnitude faster (!) on the same platform.

\begin{figure}
 \includegraphics[width=1.0\columnwidth]{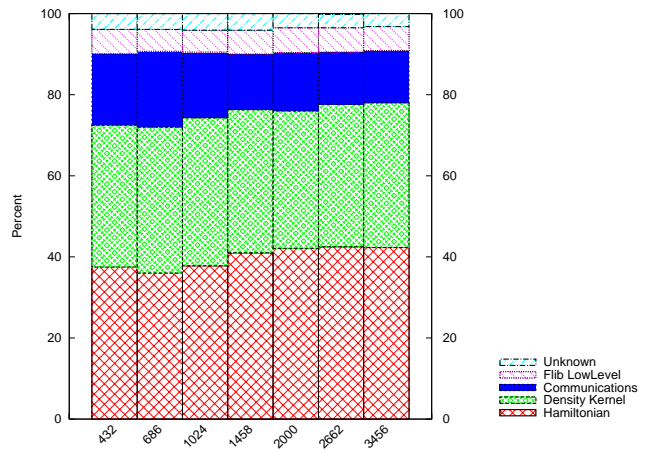}
 \caption{Portions of the overall walltime spent in the different sections of the code
 for some of the FOE runs of Fig.~\ref{fig:scaling_comparison}.
 The overall behavior of the code is similar over a wide range of system sizes, indicated by the number of atoms on the $x$-axis.
 The FOE operations to construct the density kernel take a little less than 40\% of the overall walltime.}
 \label{fig:hists}
\end{figure}

Nonetheless, this behavior is not related to the absence of a gap, 
but is due to the the unbiased nature of the description, which requires support functions optimized \emph{in-situ}. 
The reasons for this claim are explained in the following.
In Fig.~\ref{fig:hists} we show the
percentage of the time spent in the different sections of the code for the FOE runs of 
Fig.~\ref{fig:scaling_comparison}. We see that about 40\% of the time is spent in
the application of the KS Hamiltonian,
40\% in the determination of the density matrix
and some 20\% in communications, and this over a wide range of number of atoms.
All these calculations converged in about 15 iterations of the combined self-consistent optimization of the support functions and the density kernel.
This fact shows, on the one hand, that FOE calculations of the kernel are not necessarily a 
bottleneck, even when higher polynomial degrees are needed,
i.e.\ systems without a gap can be calculated efficiently with our method.
On the other hand, it will be of much interest to 
work with \emph{pre-optimized} basis functions that exhibit a similar accuracy,
in the same spirit as the fragment-based approach that has been employed using molecular fragments 
with BigDFT \cite{Ratcliff2015Fragment,Ratcliff2015}; this would provide results in only one (instead of 15) iterations
and therefore lead to calculations running more than one order of magnitude faster.
Work is ongoing in this direction.

% \subsubsection{Parallel scaling}
% We have demonstrated that \textsc{BigDFT} is capable of performing accurate linear scaling calculations for large metallic systems.
% However, even if the scaling is eventually linear, the prefactor is rather large, 
% which is a consequence of the challenges related to metallic systems discussed in Sec.~\ref{sec:Challenges for metallic systems}.
% Therefore, it is important that the code exhibits a good parallel scalability, 
% such that the walltime can be reduced to values usable in practice.
% 
% In Fig.~\ref{fig:parallel_scaling} we show the parallel scaling of \textsc{BigDFT} as a function of the number of cores, ranging from 1440 cores (the smallest possible setup due to memory limitations) up to 7680 cores.
% As test a system we use the 8x8x8 supercell, containing 1024 atoms.
% Even though this system is not extremely large and the first data point already uses a rather large number of cores, we see that the code keeps scaling well.
% \begin{figure}
%  \includegraphics[width=1.0\columnwidth]{parallel_scaling.eps}
%  \caption{Parallel scaling of \textsc{BigDFT} for the 8x8x8 supercell, containing 1024 atoms. The runs were perform using 8 OpenMP threads, and only the number of MPI tasks was varied. The speedup is shown with respect to the first data point.
%  }
%  \label{fig:parallel_scaling}
% \end{figure}

\section{Conclusions and outlook}
% Large scale DFT calculations for metallic systems are a challenging field.
% Apart from the intrinsic difficulties of metals --- which render also traditional cubic scaling DFT calculations difficult --- there are further obstacles for linear scaling approaches.
% Actually, most of the linear scaling approaches were designed for insulators, and their straightforward application to metals is questionable; even if they eventually lead to correct results, their performance will in general be limited.

In this work we have demonstrated, by applying our code \textsc{BigDFT} to the case of large Tungsten systems, that it is possible to perform accurate and efficient linear scaling DFT calculations for metals with this code.
Even though the linear scaling version of \textsc{BigDFT} was designed --- as most other codes --- for insulating systems, 
the obtained performance --- considering also the very high accuracy of the resulting description --- for metallic systems is excellent.
We have shown that the results obtained with the linear scaling version of \textsc{BigDFT} are of equal quality as those obtained with traditional cubic scaling approaches,
but the reduced scaling allows to tackle much larger systems.
The crossover point between the cubic and linear scaling treatment lies at about 500 atoms.

Thanks to these achievements, the possibility of addressing the challenge of unbiased first-principles investigation for
systems with such a large degree of complexity opens up new interesting opportunities, as now
more realistic conditions, like for instance lower concentrations of supercell defects, can be considered.
Nevertheless it has to be pointed out that, despite the very good performance offered by our approach, calculations like the one presented remain extremely challenging from a first-principles point of view.
Therefore, exploiting the full potential of quantum-mechanical investigations of metallic systems at such sizes
will only be possible if they are considered as \emph{complementary} 
investigation techniques alongside other approaches at this scale.

% \SM{Write here something about the implication this has for Fusion Reactor Research... Mar\'ia Jos\'e could you again do this?}

\section{Acknowledgments}
We acknowledge valuable discussions with Mar\'ia Jos\'e Caturla and Chu-Chun Fu.
S.M. acknowledges support from the MaX project, which has received funding from the European Union’s Horizon 2020 Research and Innovation Programme under Grant Agreements 676598.
M.A. acknowledges support from the Novartis Universit{\"a}t Basel Excellence Scholarship for Life Sciences and the Swiss National Science Foundation (P300P2-158407, P300P2-174475).
We gratefully acknowledge the computing resources on Marconi-Fusion under the EUROfusion project BigDFT4F, from the Swiss National Supercomputing Center in Lugano (project s700), the Extreme Science and Engineering Discovery Environment (XSEDE) (which is supported by National Science Foundation grant number OCI-1053575), the Bridges system at the Pittsburgh Supercomputing Center (PSC) (which is supported by NSF award number ACI-1445606), the Quest high performance computing facility at Northwestern University, and the National Energy Research Scientific Computing Center (DOE: DE-AC02-05CH11231).
% \SM{Write here if you want to acknowledge something.}

\bibliography{citationlist}

\end{document}